\documentclass[conference]{IEEEtran}
\IEEEoverridecommandlockouts
\usepackage{authblk}
\usepackage{varwidth}
\usepackage{graphicx}
\usepackage{xspace}
\usepackage{xcolor}
\usepackage{comment}
\usepackage{geometry}
\usepackage{amsmath}
\usepackage{hyperref}
\usepackage{multirow}
\usepackage{multicol}
\usepackage{color}
\usepackage{xcolor}
\usepackage{colortbl}
\usepackage{booktabs}
\usepackage{flushend}
\usepackage{booktabs}
\usepackage{tabularx}
\usepackage{caption}
\usepackage{booktabs}

\renewcommand{\IEEEtitletopspaceextra}{\dimexpr-\headheight-\headsep+18pt\relax}

\usepackage{cite}
\usepackage{amsmath,amssymb,amsfonts}
\usepackage{algorithmic}
\usepackage{graphicx}
\usepackage{textcomp}
\usepackage{xcolor}
\def\BibTeX{{\rm B\kern-.05em{\sc i\kern-.025em b}\kern-.08em
    T\kern-.1667em\lower.7ex\hbox{E}\kern-.125emX}}
\begin{document}

\title{Measuring Pain in Sickle Cell Disease using Clinical Text\\
\thanks{This paper is based on work supported by the National Institutes of
Health under Grant no 1R01AT010413-01. Any opinions, findings, and
conclusions or recommendations expressed in this material are those of the
authors and do not necessarily reflect the views of the NIH. \
The experimental procedures involving human subjects described in this paper were approved by the Institutional Review Board.}
}



\author[1]{Amanuel Alambo\thanks{Corresponding author: Amanuel Alambo (\textcolor{blue}{alambo.2@wright.edu})}}

\author[1]{Ryan Andrew}
\author[2]{Sid Gollarahalli}
\author[2]{Jacqueline Vaughn}
\author[1]{\\Tanvi Banerjee}
\author[1]{Krishnaprasad Thirunarayan}
\author[3]{Daniel Abrams}
\author[2]{Nirmish Shah}


\affil[1]{Wright State University} \affil[2]{Duke University} \affil[3]{Northwestern University}

\maketitle

\begin{abstract}
Sickle Cell Disease (SCD) is a hereditary disorder of red blood cells in humans. Complications such as pain, stroke, and organ failure occur in SCD as malformed, sickled red blood cells passing through small blood vessels get trapped. Particularly, acute pain is known to be the primary symptom of SCD. The insidious and subjective nature of SCD pain leads to challenges in pain assessment among Medical Practitioners (MPs). Thus, accurate identification of markers of pain in patients with SCD is crucial for pain management. Classifying clinical notes of patients with SCD based on their pain level enables MPs to give appropriate treatment. We propose a binary classification model to predict pain relevance of clinical notes and a multiclass classification model to predict pain level. While our four binary machine learning (ML) classifiers are comparable in their performance, Decision Trees had the best performance for the multiclass classification task achieving 0.70 in F-measure. Our results show the potential clinical text analysis and machine learning offer to pain management in sickle cell patients.
\end{abstract}


\section{Introduction}

Sickle cell disease (SCD) affects nearly 100,000 people in the US\footnote{\url{https://www.hematology.org/Patients/Anemia/Sickle-Cell.aspx}} and is an inherited red blood cell disorder. Common complications of SCD include acute pain, organ failure, and early death \cite{mohammed2019machine}. Acute pain arises in patients when blood vessels are obstructed by sickle-shaped red blood cells mitigating the flow of oxygen, a phenomenon called vaso-occlusive crisis. Further, pain is the leading cause of hospitalizations and emergency department admissions for patients with SCD. The numerous health care visits lead to a massive amount of electronic health record (EHR) data, which can be leveraged to investigate the relationships between SCD and pain. Since SCD is associated with several complications, it is important to identify clinical notes with signs of pain from those without pain. It is equally important to gauge changes in pain for proper treatment.

Due to their noisy nature, analyzing clinical notes is a challenging task. In this study, we propose techniques employing natural language processing, text mining and machine learning to predict \textbf{pain relevance} and \textbf{pain change} from SCD clinical notes.  We build two kinds of models: 1) A binary classification model for classifying clinical notes into \textit{pain relevant} or \textit{pain irrelevant}; and 2) A multiclass classification model for classifying the \textit{pain relevant} clinical notes into i) \textit{pain increase}, ii) \textit{pain uncertain}, iii) \textit{pain unchanged}, and iv) \textit{pain decrease}. We experiment with Logistic Regression, Decision Trees, Random Forest, and Feed Forward Neural Network (FFNN) for both the binary and multiclass classification tasks. For the multiclass classification task, we conduct ordinal classification as the task is to predict pain change levels ranging from \textit{pain increase} to \textit{pain decrease}. We evaluate the performance of our ordinal classification model using graded evaluation metrics proposed in \cite{gaur2019knowledge}.

\section{Related Work}

There is an increasing body of work assessing complications within SCD. Mohammed et al. \cite{mohammed2019machine} developed an ML model to predict early onset organ failure using physiological data of patients with SCD. They used five physiologic markers as features to build a model using a random forest classifier, achieving the best mean accuracy in predicting organ failure within six hours before the incident. Jonassaint et al. \cite{jonassaint2015usability} developed a mobile app to monitor signals such as clinical symptoms, pain intensity, location and perceived severity to actively monitor pain in patients with SCD. Yang et al. \cite{yang2018improving} employed ML techniques to predict pain from objective vital signs shedding light on how objective measures could be used for predicting pain.

Past work on predicting pain or other comorbidities of SCD, has thus, relied on features such as physiological data to assess pain for a patient with SCD. In this study, we employ purely textual data to assess the prevalence of pain in patients and whether pain increases, decreases or stays constant. 

There have been studies on clinical text analysis for other classification tasks. Wang et al. \cite{wang2019clinical} conducted smoking status and proximal femur fracture classification using the i2b2 2006 dataset. Chodey et al. \cite{chodey2016clinical} used ML techniques for named entity recognition and normalization tasks. Elhadad et al. \cite{elhadad2015semeval} conducted clinical disorder identification using named entity recognition and template slot filling from the ShARe corpus (Pradhan et al., 2015) \cite{pradhan2014evaluating}. Similarly, clinical text can be used for predicting the prevalence and degree of pain in sickle cell patients as it has a rich set of indicators for pain.

\section{Data Collection}

Our dataset consists of 424 clinical notes of 40 patients collected by Duke University Medical Center over two years (2017 - 2019). The clinical notes are jointly annotated by two co-author domain experts. There are two rounds of annotation conducted on the dataset. In the first round, the clinical notes were annotated as \textit{relevant to pain} or \textit{irrelevant to pain}. In the second round, the \textit{relevant to pain} clinical notes were annotated to reflect \textbf{pain change}. Figure-1 shows the size of our dataset based on \textbf{pain relevance} and \textbf{pain change}. As shown, our dataset is mainly composed of \textit{pain relevant} clinical notes. Among the \textit{pain relevant} clinical notes, clinical notes labeled \textit{pain decrease} for the \textbf{pain change} class outnumber the rest. Sample \textit{pain relevant} and \textit{pain irrelevant} notes are shown in Table-I.\


\begin{table}[htbp]

\caption{Sample Clinical notes}
\begin{center}

\begin{tabular}{p{2.0cm} | p{5.5cm} }

\hline
\cline{2-2} 

\textbf{\textbf{Pain Relevance}}& \textbf{\textbf{Sample Clinical Note}} \\

\hline
YES & Patient pain increased from 8/10 to 9/10 in chest. \\

\hline
NO & Discharge home
\vspace*{-\baselineskip}




\end{tabular}
\label{tab1}
\end{center}
\vspace{-4mm}
\end{table}


Our dataset is highly imbalanced, particularly, among the \textbf{pain relevance} classes. There are significantly higher instances of clinical notes labeled \textit{pain relevant} than \textit{pain irrelevant}. To address this imbalance in our dataset, we employed a technique called Synthetic Minority Over-sampling TEchnique (SMOTE) \cite{chawla2002smote} for both classification tasks.

\begin{figure}[htbp]
\centerline{\includegraphics[width=70mm]{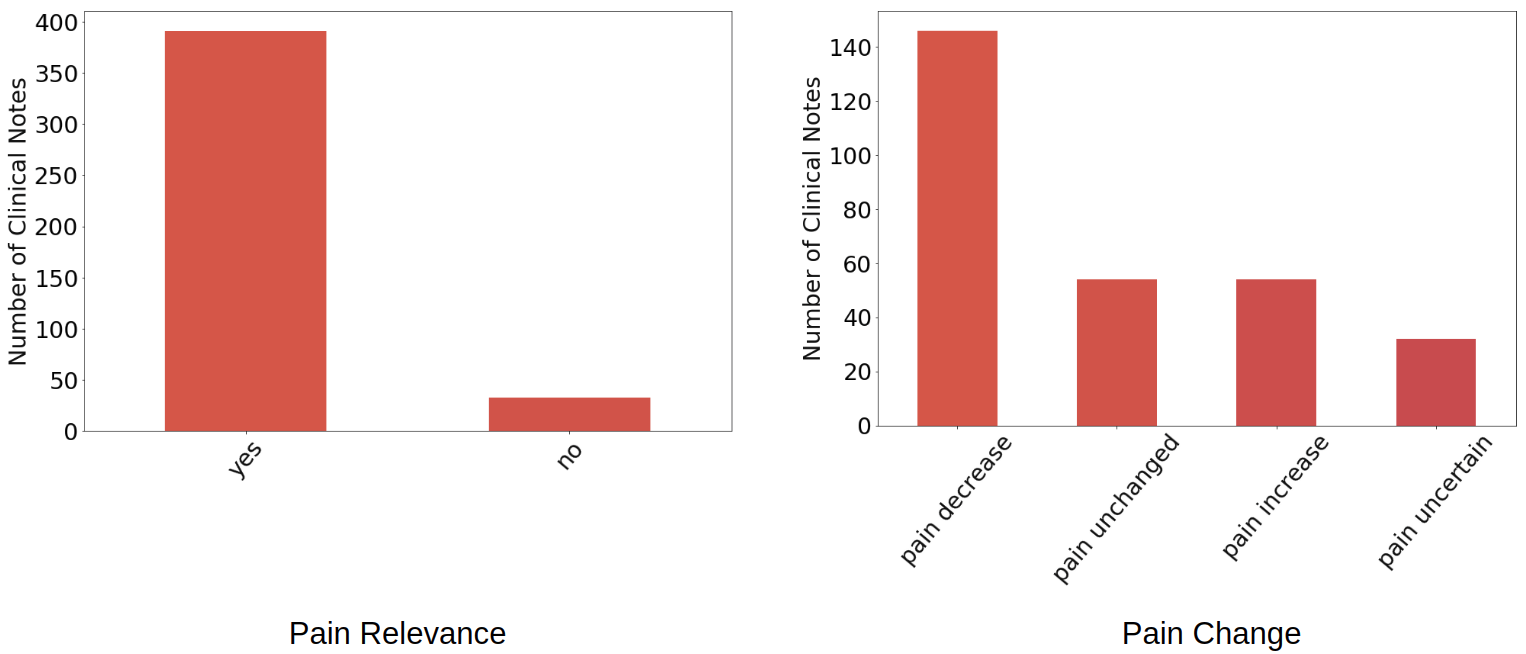}}

\caption{Statistics of dataset for \textbf{Pain Relevance} and \textbf{Pain Change} classes}
\label{fig}
\end{figure}

We preprocessed our dataset by removing stop words as well as punctuations, and performed lemmatization.

\section{Methods}

The clinical notes are labeled by co-author domain experts based on their \textbf{pain relevance} and \textbf{pain change} indicators. The \textbf{pain change} labels use a scale akin to the Likert scale from severe to mild. Our pipeline (Figure-2) consists of data collection, data preprocessing, linguistic/topical analysis, feature extraction, feature selection, model creation, and evaluation. We use linguistic and topical features to build our models. While linguistic analysis is used to extract salient features, topical features are used to mine latent features. We performed two sets of experiments: 1) Binary Classification for \textbf{pain relevance} classification, and 2) Multiclass Classification for \textbf{pain change} classification.

\begin{figure}[htbp]
\centerline{\includegraphics[width=75mm]{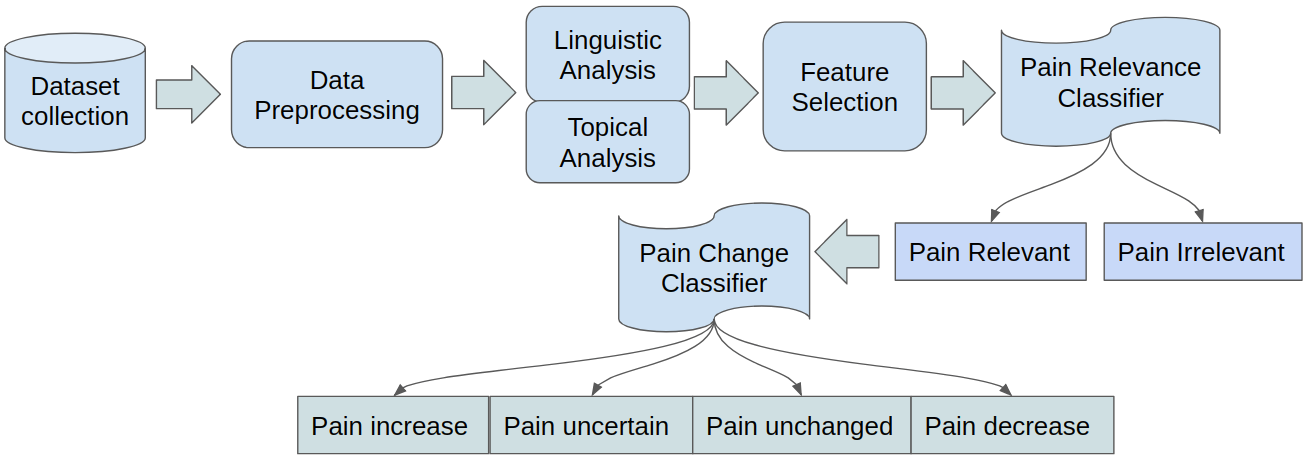}}

\caption{Sickle Cell Disease Pain Classification Pipeline}
\label{fig}
\end{figure}

\subsection{Linguistic Analysis}

To infer salient features in our dataset, we performed linguistic analysis. We generated n-grams for \textit{pain-relevant} and \textit{pain-irrelevant} clinical notes and clinical notes labeled \textit{pain increase}, \textit{pain uncertain}, \textit{pain unchanged}, or \textit{pain decrease}. In our n-grams analysis, we observe there are unigrams and bigrams that are common to different classes (e.g., common to \textit{pain relevant} and \textit{pain irrelevant}). Similarly, there are unigrams and bigrams that are exclusive to a given class. Table-II shows the top 10 unigrams selected using \( \chi^2 \) feature selection for our dataset based on the classes of interest.

\begin{table}[htbp]

\caption{Top 10 Unigrams}
\begin{center}

\begin{tabular}{p{2.15cm} |p{2cm}| p{2cm} }

\hline
\cline{2-3} 

\textbf{\textbf{Pain Relevant
(Exclusive)}}& \textbf{\textbf{Pain Irrelevant
(Exclusive)}}& \textbf{\textit{Pain Relevant AND Pain Irrelevant}} \\

\hline
emar, intervention, increase, dose, expressions, chest, regimen, alteration, toradol, medication & home, wheelchair, chc, fatigue, bedside, parent, discharge, warm, relief, mother & pain, pca, plan, develop, control, altered, patient, level, comfort, manage 



\end{tabular}
\label{tab1}
\end{center}
\vspace{-4mm}
\end{table}

\subsection{Topical Analysis}
While n-grams analysis uncovers explicit language features in the clinical notes, it is equally important to uncover the hidden features characterizing the topical distribution. We adopt the Latent Dirichlet Allocation (LDA) \cite{blei2003latent} for unraveling these latent features. We train an LDA model using our entire corpus. 

To determine the optimal number of topics for a given class of clinical notes (e.g., \textit{pain relevant} notes), we computed coherence scores \cite{stevens2012exploring}. The higher the coherence score for a given number of topics, the more intepretable the topics are (see Figure-3). We set the number of words characterizing a given topic to eight. These are words with the highest scores in the topic distribution. We found the human-interpretable optimal number of topics for each of the classes of the clinical notes in our dataset to be two. This is interpreted as each class of the clinical notes is a mixture of two topics. Table-III shows words for the two topics for \textit{pain relevant} and \textit{pain irrelevant} clinical notes. As can be seen in the table, \textit{pain relevant} notes can be interpreted to have mainly the topic of pain control, while \textit{pain irrelevant} notes to have primarily the topic of home care. Similarly, Table-IV shows the distribution of words for the topics for each of the \textbf{pain change} classes (underscored words are exclusive to the corresponding class for Topic-1). Further, \textit{pain} appears in each of the topics for \textbf{pain change} classes and, as a result, is not discriminative. While a common word such as \textit{pain} in the topic distribution can be considered as a stop word and not helpful for \textbf{pain change} classification, we did not remove it since \textit{pain} helps with interpretation of a given topic regardless of other topics.

\begin{figure}[htbp]
\centerline{\includegraphics[width=46mm, height=2.5cm]{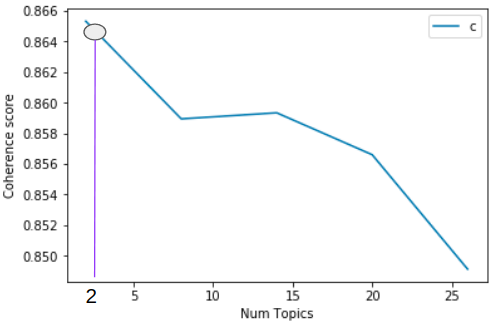}}

\caption{Coherence Scores vs Number of Topics }
\label{fig}

\vspace{-1.5em}
\end{figure}


\begin{table}[htbp]
\vspace{1 mm}
\caption{Topic distribution based on pain relevance}
\begin{center}
\vspace{-4 mm}

\begin{tabular}{p{1.25cm} | p{3.2cm} | p{3.2cm}}

\hline
\cline{2-3} 

\textbf{\textbf{Pain Relevance}}& \textbf{Most Prevalent Words in Topic-1} & \textbf{Most Prevalent Words in Topic-2} \\

\hline
YES & progress, pain, improve, decrease, knowledge, control
& patient, pain, medication, knowledge, goal, state
\\

\hline
NO & note, admission, discharge, patient, home, ability
& pain, goal, admission, outcome, relief, continue 

\vspace*{-\baselineskip}


\end{tabular}
\label{tab1}
\end{center}
\vspace{-4mm}
\end{table}

\begin{table}[htbp]

\caption{Topic distribution based on pain change}
\begin{center}

\vspace{-4 mm}

\begin{tabular}{p{1.25cm} | p{3.2cm} | p{3.2cm}}

\hline
\cline{2-2} 

\textbf{\textbf{Pain Change}}& \textbf{\textbf{Most Prevalent Words in Topic-1}} & \textbf{\textbf{Most Prevalent Words in Topic-2}}  \\

\hline
Pain increase & pain, progress, \textbf{\underline{medication}}, \textbf{\underline{management}}, patient, \textbf{\underline{schedule}}, pca,\textbf{\underline{ intervention}} & pain, patient, give, goal, intervention, dose, button, plan\\

\hline
Pain uncertain & pain, patient, \textbf{\underline{goal}}, \textbf{\underline{continue}}, plan, \textbf{\underline{improve}}, decrease, develop & outcome, pain, problem, knowledge, regimen, deficit, carry, method\\

\hline
Pain unchanged & pain, progress, \textbf{\underline{level}}, \textbf{\underline{control}}, develop, plan, regimen, pca & patient, pain, remain, well, demand, plan, level, manage\\

\hline
Pain decrease & pain, progress, patient, decrease, plan, regimen, \textbf{\underline{satisfy}}, \textbf{\underline{alter}} & pain, patient, improve, satisfy, control, decrease, manage, ability\\

\vspace*{-\baselineskip}

$^{\mathrm{}}$& &  \\

\hline

\end{tabular}
\label{tab1}
\end{center}
\vspace{-4mm}
\end{table}

\subsection{Classification}

The language and topical analyses results are used as features in building the ML models. 
Our classification task consists of two sub-classification tasks: 1) \textbf{pain relevance} classification; 2) \textbf{pain change} classification, each with its own sets of features. The \textbf{pain relevance} classifier classifies clinical notes into \textit{pain-relevant} and \textit{pain-irrelevant}. The \textbf{pain change} classifier is used to classify the \textit{pain-relevant} clinical notes into 1) \textit{pain increase}, 2) \textit{pain uncertain},  3) \textit{pain unchanged}, and 4) \textit{pain decrease}. We trained and evaluated various ML models for each classification task. We used a combination of different linguistic and topical features to train our models. Since linguistic and topical features are generated using independent underlying techniques, which make them orthogonal, concatenation operation is used to combine their representations. We split our dataset into 80\% training and 20\% testing sets and built logistic regression, decision trees, random forests, and FFNN for both classification tasks. Table-V shows the results of the \textbf{pain relevance} classifier while Table-VI shows \textbf{pain change} classification results. For the ordinal classification, we considered the following order in the severity of pain change from high to low: \textit{pain increase}, \textit{pain uncertain}, \textit{pain unchanged}, \textit{pain decrease}.

\begin{table}[ht]
    \caption{Pain Relevance Classification}
    \centering
    \scalebox{0.85}{
    \begin{tabular}{|c|c|c|c|c|}
        \hline
        \textbf{Model} & \textbf{Feature} & \textbf{Precision} & \textbf{Recall} & \textbf{F-measure}\\
        \hline
        \multirow{3}{*} {Logistic Regression} & Linguistic & 0.94 & 0.93 & 0.94\\
                        & Topical & 0.98 & 0.86 & 0.91\\
                        & Linguistic + Topical & 0.95 & 0.95 & 0.95\\
        \hline
        \multirow{3}{*} \textbf{Decision Trees} & Linguistic & 0.95 & 0.95 & 0.95\\
                        & Topical & 0.98 & 0.98 & 0.98\\
                        & \textbf{Linguistic + Topical} & \textbf{0.98} & \textbf{0.98} & \textbf{0.98}\\
        \hline
        
        \multirow{3}{*} {Random Forest} & Linguistic & 0.90 & 0.95 & 0.92\\
                        & Topical & 0.95 & 0.98 & 0.98\\
                        & Linguistic + Topical & 0.90 & 0.95 & 0.93\\

        \hline
        \multirow{3}{*} {FFNN} & Linguistic & 0.94 & 0.94 & 0.94\\
                        & Topical & 0.98 & 0.98 & 0.98\\
                        & Linguistic + Topical & 0.96 & 0.96 & 0.94\\
        \hline
    \end{tabular}
    }
    \label{multirow_table}
\end {table}

\begin{table}[ht]
    \caption{Pain Change Classification}
    \centering
    \scalebox{0.85}{
    \begin{tabular}{|c|c|c|c|c|}
        \hline
        \textbf{Model} & \textbf{Feature} & \textbf{Precision} & \textbf{Recall} & \textbf{F-measure}\\
        \hline
        \multirow{3}{*} {Logistic Regression} & Linguistic & 0.75 & 0.56 & 0.63\\
                        & Topical & 0.50 & 0.55 & 0.52\\
                        & Linguistic + Topical & 0.76 & 0.58 & 0.66\\
        \hline
        \multirow{3}{*} \textbf{Decision Trees} & Linguistic & 0.76 & 0.59 & 0.67\\
                        & Topical & 0.73 & 0.65 & 0.68\\
                        & \textbf{Linguistic + Topical} & \textbf{0.74} & \textbf{0.68} & \textbf{0.70}\\
        \hline
        
        \hline
        \multirow{3}{*} {Random Forest} & Linguistic & 0.74 & 0.49 & 0.59\\
                        & Topical & 0.94 & 0.52 & 0.66\\
                        & Linguistic + Topical & 0.81 & 0.46 & 0.59\\

        \hline
        \multirow{3}{*} {FFNN} & Linguistic & 0.71 & 0.59 & 0.65\\
                        & Topical & 0.73 & 0.65 & 0.68\\
                        & Linguistic + Topical & 0.83 & 0.51 & 0.63\\
        \hline
    \end{tabular}}
    \label{multirow_table}

\end {table}
                    
\section{Discussion}

For \textbf{pain relevance} classification, the four models have similar performance. For \textbf{pain change} classification, however, we see a significant difference in performance across the various combinations of features and models. Decision trees with linguistic and topical features achieve the best performance in F-measure. While random forest, and FFNN offer better precision, each, than decision tree, they suffer on Recall, and therefore on F-measure. Further, most models perform better when trained on topical features than pure linguistic features. A combination of topical and linguistic features usually offers the best model performance. Thus, latent features obtained using LDA enable an ML model to perform better. 

Evaluation of the multiclass classification task is conducted using the techniques used by Gaur et al. \cite{gaur2019knowledge}  where a model is penalized based on how much it deviates from the true label for an instance. Formally, the count of true positives is incremented when the true label and predicted label of an instance are the same. Similarly, false positives’ count gets incremented by an amount equal to the gap between a predicted label and true label (when predicted label is higher than true label). False negatives’ count is incremented by the difference between the predicted label and true label (when predicted label is lower than true label). Precision, and recall are then computed following the implementations defined in ML libraries\footnote{\url{https://bit.ly/3a5Fibb}} using the count of true positives, false positives, and false negatives. Finally, F-measure is defined as the harmonic mean of precision and recall.

While we achieved scores on the order of 0.9 for \textbf{pain relevance} classification, the best we achieved for \textbf{pain change} classification was 0.7. This is because there is more disparity in linguistic and topical features between \textit{pain relevant} and \textit{pain irrelevant} notes than there is among the four \textbf{pain change} classes. Since the price of false negatives is higher than false positives in a clinical setting, we favor decision trees with n-grams and topics used as features as they achieve the best Recall and F-measure, albeit they lose to other models on Precision. Thus, identification of \textit{pain relevant} notes with 0.98 F-measure followed by a 0.70 F-measure on determining \textit{pain change} is impressive. We believe our model can be used by MPs for SCD-induced pain mitigation.

\vspace{8mm}

\section{Conclusion and Future Work}

In this study, we conducted a series of analyses and experiments to leverage the power of natural language processing and ML to predict \textbf{pain relevance} and \textbf{pain change} from clinical text. Specifically, we used a combination of linguistic and topical features to build different models and compared their performance. Results show decision tree followed by feed forward neural network as the most promising models.

In future work, we plan to collect additional clinical notes and use unsupervised, and deep learning techniques for predicting pain. Further, we look forward to fusing different modalities of sickle cell data for better modeling of pain or different physiological manifestations of SCD.



\vspace{12pt}

\bibliographystyle{IEEEtran}  
\bibliography{bibliography.bib}

\end{document}